\documentclass[aps,prl,twocolumn,groupedaddress,fleqn]{revtex4}
\pdfoutput=1
\usepackage{graphicx}
\usepackage{amsmath}
\DeclareMathOperator{\sn}{sn}
\DeclareMathOperator{\cn}{cn}
\DeclareMathOperator{\dn}{dn}

\begin{document}

\title{Stable equilibrium point and oscillatory motion of the Universe in a model with variable vacuum energy}
\author{K. Trachenko$^{1}$}
\address{$^1$ School of Physics and Astronomy, Queen Mary University of London, Mile End Road, London, E1 4NS, UK}

\begin{abstract}
We discuss the mechanism by which the field vacuum energy varies as a result of strong self-interaction. We propose a non-perturbative approach to treat strong interactions and discuss the problem in terms of quasi-particles describing the motion of field modes. The resulting vacuum energy is variable and depends on the state of the system. If the interacting scalar field is related to the cosmological field, an acceleration equation with a stable equilibrium point follows in a simple model, predicting the oscillatory behavior of the scale factor and other cosmological effects.
\end{abstract}

\maketitle

\section{Introduction}

The zero-point, or vacuum, energy is an interesting property emerging in quantum mechanics and playing an important role in other areas. The operation of the vacuum energy is not well understood. A major challenge is seen in uncovering an underlying physical structure that governs the vacuum energy and its evolution \cite{carroll,adams,liddle,linde,roos,bass,silve,sahni,weinberg,sola-review}. More specifically, the question is what mechanism can alter the vacuum energy of the field, with the scalar field being a commonly considered case \cite{carroll,adams,liddle,linde,roos}?

A useful insight comes from recent results in condensed matter physics. In solids, the vacuum energy is the sum of zero-point energies of collective modes, phonons, and is viewed to form a constant energy background. Strongly-interacting and dynamically disordered liquids have been much harder to understand because inter-particle interactions are strong and the perturbation theory does not apply \cite{landau}. A {\it non-perturbative} approach to liquid thermodynamics \cite{ropp} gives the important result that solid-like phonons in liquids evolve: the number of high-frequency solid-like transverse modes reduces with temperature. As a result, the liquid vacuum energy becomes a variable property that depends on the state of the system \cite{jpcm}. At some high temperature, transverse modes are lost at all available frequencies and the vacuum energy of liquid transverse modes becomes zero.

The collective modes in condensed matter are equivalent to field modes in the field theory \cite{bogol} where the vacuum energy is the sum of energies of field modes. This opens an interesting possibility that the mechanism that varies the zero-point energy in liquids can apply to the vacuum energy of fields. We have recently started to explore this question and suggested that this would be the case if the field Hamiltonian includes self-interaction similar to the double-well potential in liquids \cite{ropp,annals}.

In this paper, we discuss the variability of the vacuum energy as a result of strong self-interaction of the field. In the next section, we review the mechanism by which the vacuum energy varies in liquids as a result of strong anharmonic interactions. We then consider strongly-interacting fields and propose a non-perturbative approach based on the concept of quasi-particles. This gives variable vacuum energy that depends on the state of the system. We consider the field to be a cosmological scalar field related to inflation and discuss a simple model which yields an acceleration equation with a stable equilibrium point. The model predicts the oscillatory behavior of the scale factor, and we discuss other cosmological implications including the cosmological constant problem.

\section{Variable vacuum energy}

\subsection{Evolution of collective modes in liquids}

In solids, interactions are commonly treated in the harmonic approximation with a Hamiltonian

\begin{equation}
H_s(x)=\frac{1}{2}\sum_im_i\dot{x_i}^2+\frac{1}{2}\sum_{ij}k_{ij}(x_i-x_j)^2
\label{sol}
\end{equation}

\noindent where $x_i$ and $m_i$ are particle coordinates and masses and $k_{ij}$ are stiffness coefficients.

Eq. (\ref{sol}) can be transformed to the diagonal form in terms of normal coordinates $u_i$

\begin{equation}
H_s(u)=\frac{1}{2}\sum_i\left(\dot{u_i}^2+\omega_i^2 u_i^2\right)
\label{sol1}
\end{equation}

\noindent where $\omega_i$ are normal frequencies.

Eq. (\ref{sol1}) is equivalent to the Hamiltonian of the free field discussed in the next section.

\begin{figure}
\begin{center}
{\scalebox{0.33}{\includegraphics{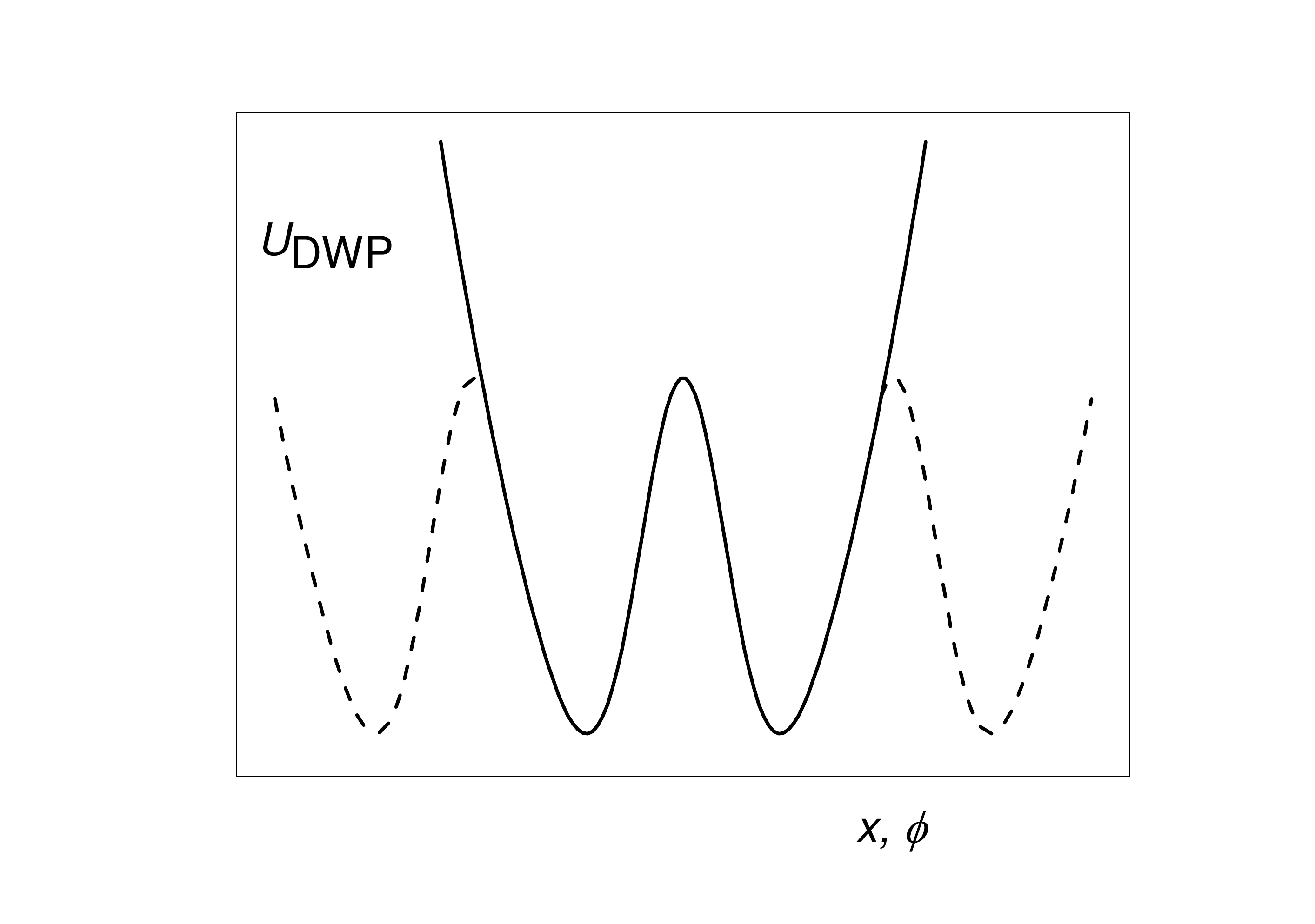}}}
\end{center}
\caption{Schematic representation of the double-well potential $U_{\rm DWP}(x_i)$.}
\label{wells}
\end{figure}

In liquids, one needs to modify (\ref{sol}) in order to endow the particles with the ability to jump between quasi-equilibrium positions. This requires adding non-linear terms to the interaction term. The non-linear terms are not small and so the perturbation theory does not apply: the perturbation approach applies to weakly-interacting gases only but not to strongly-interacting liquids. The ensuing treatment becomes intractable because it involves a large number of coupled non-linear oscillators, an exponentially complex problem \cite{ropp}. Impossible to solve mathematically, the problem becomes amenable to treatment using a physical approach and recognizing that the dynamics of particles in liquids consists of two types: solid-like oscillatory motion around quasi-equilibrium positions and jumps between different positions that enable liquid flow \cite{frenkel}. This corresponds to motion in the double-well (or multi-well) potential shown in Figure \ref{wells}: particles oscillate in one potential minimum and jump between different minima. The precise form of the potential is non-essential; only the existence of two minima separated by a finite energy barrier is.

The key parameter in this consideration is liquid relaxation time $\tau$ introduced by J Frenkel, the average time between particle jumps from one potential minimum to the next \cite{frenkel}. $\tau$ depends on the height of the energy barrier in Figure \ref{wells}. We denote the inverse of $\tau$ as $\omega_{\rm F}$, the Frenkel hopping frequency. It now follows that at times shorter than $\tau$ or frequencies above $\omega_{\rm F}$, the system is a solid and therefore supports two solid-like transverse modes with frequency above $\omega_{\rm F}$. At times longer than $\tau$ or frequencies below $\omega_{\rm F}$, the system is a flowing liquid capable of supporting density fluctuations only (as any elastic medium) but not solid-like transverse modes. This is the main effect of how particle jumps due to non-linear terms in (\ref{sol}) modify collective modes in liquids \cite{ropp}.

Therefore, the Hamiltonian describing the energy of transverse phonons in liquids reads as:

\begin{equation}
H(u)=\frac{1}{2}\sum_{\omega_i>\omega_{\rm F}}\left(\dot{u_i}^2+\omega_i^2 u_i^2\right)
\label{liquid}
\end{equation}

\noindent where $u$ is the normal phonon coordinate.

The energy corresponding to (\ref{liquid}), $E_t$, is:

\begin{equation}
E_t=\int\limits_{\omega_{\rm F}}^{\omega_{\rm D}}E(\omega,T)g_t(\omega)d\omega
\label{l1}
\end{equation}

\noindent where $\omega_{\rm D}$ is Debye frequency of transverse modes close to the maximal frequency in the system, $g_t(\omega)=\frac{6N}{\omega_{\rm D}^3}\omega^2$ is Debye density of states for transverse modes and $E(\omega,T)$ is the phonon energy.

In the classical case, $E=T$ (here and below, $k_{\rm B}=1$), and integrating (\ref{l1}) gives

\begin{equation}
E_t=2NT\left(1-\left(\frac{\omega_{\rm F}}{\omega_{\rm D}}\right)^3\right)
\label{l2}
\end{equation}

\noindent and agrees with experimental specific heats of many liquids \cite{ropp}.

More relevant for this discussion, the zero-point energy of transverse modes in the liquid, $E_0$, can be calculated using $E(\omega,T)=\frac{\hbar\omega}{2}$ in (\ref{l1}) and gives \cite{jpcm}:

\begin{equation}
E_0=\frac{3N}{4}\hbar\omega_{\rm D}\left(1-\left(\frac{\omega_{\rm F}}{\omega_{\rm D}}\right)^4\right)
\label{l3}
\end{equation}

The hopping frequency $\omega_{\rm F}$ increases with temperature and decreases with pressure. Therefore, $E_0$ in (\ref{l3}) becomes variable. When $\omega_{\rm F}$ reaches its limiting value of $\omega_{\rm D}$, transverse modes disappear completely, and $E_0=0$.

In terms of particle dynamics, $\omega_{\rm F}\rightarrow\omega_{\rm D}$ corresponds to the Frenkel line on the phase diagram we introduced recently \cite{prl,phystoday}. The FL corresponds to the transition of particle motion from the combined oscillatory and diffusive below the line to purely gas-like diffusive above the line. In terms of the potential in Figure \ref{wells}, particles oscillate and jump below the FL and move diffusively above the activation barrier above the FL. Crossing the FL has important implications for the phonon states: two solid-like transverse modes disappear completely at the line and the remaining longitudinal mode starts disappearing starting from the maximal frequency above the line \cite{ropp}.

Notably, the above calculation of the liquid energy in (\ref{l2}) and (\ref{l3}) is done in a {\it non-perturbative} approach based on the introduction of $\tau$ ($\omega_{\rm F}$). Indeed, a perturbation theory assumes only small deviations from the harmonic minimum in Figure 1 and hence is unable to describe the hopping regime between different wells in the liquid. 

As far as the liquid Hamiltonian is concerned, introducing particle jumps between quasi-equilibrium positions with a certain frequency is equivalent to introducing an anharmonic term $V(x)$ resulting in a double-well (or multi-well) potential in Figure \ref{wells} which enables particles to jump. This gives the liquid Hamiltonian as

\begin{equation}
H(x)=\sum\limits_i\frac{1}{2}m_i\dot{x_i}^2+\frac{1}{2}\sum\limits_{ij}k_{ij}(x_i-x_j)^2+\sum\limits_i V(x_i)
\label{dwp0}
\end{equation}

\noindent where the anharmonic term $V(x)$ gives the second well separated by the potential energy barrier. The height of the barrier is set by the energy of particle interaction.

Eq. (\ref{dwp0}) can be re-written as

\begin{eqnarray}
&H(x)=\sum\limits_i\left(\frac{1}{2}m_i\dot{x_i}^2+U_{\rm DWP}(x_i)\right)\nonumber\\
&U_{\rm DWP}(x_i)=\frac{1}{2}\sum\limits_{j}k_{ij}(x_i-x_j)^2+V(x_i)
\label{dwp}
\end{eqnarray}

\noindent where $U_{\rm DWP}(x_i)$ is the double-well potential in Figure \ref{wells}.

We have explained how $U_{\rm DWP}(x_i)$ originates in liquids, however the potential in Figure \ref{wells} is a general construction used to discuss three different regimes of particle dynamics and three states of matter: solids, liquids and gases. Particles oscillate in one single minimum in solids, oscillate and diffusively move between different minima in liquids and diffusively move above the potential barrier in gases. If $K$ and $P$ are kinetic and potential energy of the system, $K\ll P$, $K\approx P$ and $K\gg P$ give solid, liquid and gas state, respectively.

\subsection{Interacting fields and quasiparticles}

We consider a self-interacting scalar field $\phi$ at a finite temperature and discuss the evolution of the field energy as a result of self-interaction. The field Hamiltonian is

\begin{eqnarray}
&H(\phi)=\frac{1}{2}\sum\limits_i\left(\dot{\phi_i}^2+\omega_i^2\phi_i^2+V(\phi_i)\right)\nonumber \\
&U_{\rm DWP}(\phi_i)=\omega_i^2\phi_i^2+V(\phi_i)
\label{field}
\end{eqnarray}

In (\ref{field}), the first two terms represent independent harmonic modes with frequencies $\omega_i$ as is the case for, for example, Klein-Gordon field. The non-interacting Hamiltonian has positive $\omega_i^2$. $V(\phi)$ in (\ref{field}) is the interaction term, and we assume that $U_{\rm DWP}(\phi_i)=\omega_i^2\phi_i^2+V(\phi_i)$ has a double-well form in Figure \ref{wells} (the same that gives rise to three states of matter), endowing the field with the ability to oscillate around different quasi-equilibrium values. This general form is the only assumption in the theory; precise form of $V(\phi)$ is unimportant. Our main qualitative results related to the vacuum energy follow without further assumptions.

It is interesting to ask about the origin of the double-well (or multi-well) potential in Figure \ref{wells} for the field and how generic it is. One motivation comes from the form of the Higgs field. We also note that commonly considered interaction terms in the field theory (e.g., $\phi^3$, $\phi^4$ and so on) are often chosen for their computational tractability and renormalizability rather than on specific physical grounds. Our interest in the double-well form $V(\phi_i)$ is that it gives different regimes of field dynamics as it does for particle dynamics.

We now interpret the field mode coordinate $\phi_i$ in (\ref{field}) as the coordinate of a particle moving in a double-well potential in Figure \ref{wells} at a given temperature. In other words, we consider variables $\phi_i$ as {\it quasiparticles}. Then, (\ref{field}) describes quasiparticles moving in a double-well potential. As in the case of liquid particles, this motion consists of solid-like oscillatory motion in a single well in Figure \ref{wells} (corresponding to the free field) and diffusive hopping between the wells with a certain frequency $\omega_{\rm F}$ or time $\tau$ ($\omega_{\rm F}=\frac{1}{\tau}$) that depends on the barrier height. This implies that the oscillatory motion of quasiparticles in each well can take place only above frequency $\omega_{\rm F}$ but not below (the period of quasiparticle oscillation in one well can not exceed the hopping time $\tau$). In other words, the interaction results in the emergence of the frequency gap $\omega_{\rm F}$, or energy gap $\hbar\omega_{\rm F}$.

Hence, similarly to liquids, the anharmonic double-well interaction modifies the spectrum of the free field by introducing the lower frequency cutoff, $\omega_{\rm F}$, and the Hamiltonian becomes (compare to (\ref{liquid})):

\begin{equation}
H(\phi)=\frac{1}{2}\sum_{\omega_i>\omega_{\rm F}}\left(\dot{\phi_i}^2+\omega_i^2\phi_i^2\right)
\label{field1}
\end{equation}

We note that the interacting field Hamiltonian (\ref{field}) is not identical to its particle counterpart (\ref{dwp}). The field Hamiltonian (\ref{field}) is simpler because all its terms are diagonal. The quadratic terms in (\ref{dwp}) can be diagonalized as in (\ref{sol1}) but the this gives cross terms in $V(x)$. However, this circumstance does not change the character of motion of quasiparticles $\phi$: the double-well potential in (\ref{field}) implies that the motion of quasiparticles $\phi_i$ consists of oscillations in a single well as in the case of the free field and the hopping motion between different wells.

The zero-point (vacuum) energy corresponding to (\ref{field1}) can be evaluated as:

\begin{equation}
E_0=\int\limits_{\omega_{\rm F}}^{\omega_{\rm max}}\frac{\hbar\omega}{2} g(\omega)d\omega
\label{field2}
\end{equation}

\noindent where $\omega_{\rm max}$ is the maximal frequency. We do not specify $\omega_{\rm max}$ for now and view it as a model parameter.

We consider a general case of $n$-component scalar field. This gives the density of states per mode as $g(\omega)=\frac{n^2}{\omega_{\rm max}^n}\omega^{n-1}$. As written, $g(\omega)$ ensures
that the total number of modes divided by the number of modes in each component is $n$. Integrating ({\ref{field2}) gives

\begin{equation}
E_0=\frac{n^2}{2(n+1)}\hbar\omega_{\rm max}\left(1-\left(\frac{\omega_{\rm F}}{\omega_{\rm max}}\right)^{n+1}\right)
\label{zero}
\end{equation}

Eq. (\ref{zero}) describes the mechanism of variation of the vacuum energy and is an important independent result in this paper. In (\ref{zero}), the vacuum energy changes with $\omega_{\rm F}$ which, in turn, changes with external parameters. When $\omega_{\rm F}\rightarrow\omega_{\rm max}$, $E_0$ tends to zero. In liquids, this corresponds to the Frenkel line separating the
states with and without propagating transverse modes as discussed earlier.

We note that Eq. (\ref{zero}) can not be derived in a perturbation theory because the anharmonic interaction term is not small. Indeed, the interaction term results in hopping of field quasiparticles between different minima, the effect which can not operate if the displacement from the minimum in the harmonic non-interacting potential is small as assumed in the perturbation theory. Therefore, our quasiparticle approach to calculating the vacuum energy of strongly-interacting field is {\it non-perturbative}.

Before proceeding further, we make two remarks. First, the variation of the vacuum energy does not need to involve a phase transition. In liquids, phonon states change qualitatively at the first-order liquid-gas phase transition line. However, the important change of phonon states can also take place above the critical point where no phase transition operates. This takes place at the Frenkel line \cite{prl,phystoday,ropp} discussed earlier: crossing the FL corresponds to the complete loss of solid-like transverse modes in the spectrum. Similarly, the variation of the field vacuum energy in (\ref{zero}) does not need to involve a phase transition and can operate above the critical point. In this case, issues related to the nucleation of phases, boundaries and other effects \cite{liddle,linde} do not emerge.

Considering field dynamics in the supercritical state can be thought to be more natural as compared to the first-order phase transition. Indeed, the Frenkel line extends to arbitrarily high temperature and pressure above the critical point on the phase diagram \cite{ropp}, in contrast to the liquid-gas transition line which is bound between the triple and critical points. Hence, the range of parameters on the phase diagram that give the crossover in the supercritical state is infinitely larger than that corresponding to the phase transition.

Second, although the FL in liquids signifies the complete loss of two transverse modes when $\omega_{\rm F}\rightarrow\omega_{\rm max}$, the longitudinal mode remains and evolves above the line \cite{ropp}. The longitudinal mode, related to density fluctuations, exists in any medium with elastic response. One could formally extend the similarity of liquid and field Hamiltonians to include the energy of the remaining ``longitudinal'' field mode above the FL \cite{annals1}. In the quasiparticle picture proposed here, no analogue of the longitudinal mode is considered. Indeed, quasiparticles moving in the double-well potential are field mode coordinates $\phi_i$ which progressively disappear starting from $\omega_{\rm F}$. When $\omega_{\rm F}\rightarrow\omega_{\rm max}$, the field vacuum energy tends to zero.

\section{Dynamics of the scale factor in the potential with a stable equilibrium point}

As commonly discussed \cite{roos,liddle,linde}, we consider the scalar field $\phi$ discussed in the previous section to be the cosmological field driving the inflation. If $a$ is the Universe scale factor, $\rho_\Lambda$ is vacuum energy density and $P_\Lambda$ is pressure, the acceleration equation is

\begin{equation}
\frac{\ddot{a}}{a}=-\frac{4\pi G}{3c^2}\left(\rho_\Lambda+3P_\Lambda\right)
\label{2}
\end{equation}
\noindent

For now, we do not consider the contribution of matter to (\ref{2}) and will return to its effect later. We consider an exact solution to an approximate version of the fuller problem in a simple model. For the purpose of discussing the evolution of field modes, we continue to consider the field quantized in Minkowski spacetime and do not account for the Hubble damping term. The largest contribution to the vacuum energy comes from high-frequency modes with wavelengths sufficiently short not to be affected by the space curvature and to be within the horizon in the proposed model.

The vacuum energy in Eq. (\ref{zero}) depends on $\omega_{\rm F}$ which, in turn, depends on external parameters (density and temperature) and therefore on $a$. There are two factors contributing to the dependence of $\omega_{\rm F}$ on $a$: ``bare'' variation of $\omega_{\rm F}$ due to adiabatic expansion and the change of $\omega_{\rm F}$ due to redshift. The effect of the latter is readily known: $\omega_{\rm F}\propto\frac{1}{a}$ \cite{liddle}. In order to study the ``bare'' dependence of $\omega_{\rm F}$ on $a$, $\omega^{\rm bare}_{\rm F}(a)$, one needs to consider how the dynamics of quasiparticles $\phi$ in (\ref{field}) changes during the adiabatic expansion of the Universe. The precise form of $\omega^{\rm bare}_{\rm F}(a)$ depends on the potential's form in Figure \ref{wells} and is not generally known. However, we can use the quasiparticle picture (see the previous section) in which $\omega_{\rm F}$ has the same dynamical behavior for interacting fields and liquids. In liquids, $\omega_{\rm F}$ is known to increase during the adiabatic expansion because $\tau$ decreases along the adiabat. $\omega_{\rm F}$ can be calculated as $\omega_{\rm F}=\frac{1}{\tau}\propto\frac{1}{\eta}$, where we used the Maxwell relationship between $\tau$ and viscosity $\eta$, $\eta=G\tau$ ($G$ is infinite shear modulus). We use the experimental data for supercritical Ar from the National Institute of Standards and Technology database \cite{nist} and consider the state points at constant entropy because we are interested in the adiabatic expansion. We used nearly equidistant temperature points in the range 200-440 K and selected the corresponding density and $\eta$ at constant entropy. We note that a useful condition for the system to be in the hopping regime below the Frenkel line where Eqs. (\ref{liquid})-(\ref{l3}) apply is $c_v>2k_{\rm B}$, where $c_v$ is specific heat \cite{ropp}. In agreement with this condition, $c_v$ reported in the database at each state point is in the range (2.03-2.43)$k_{\rm B}$.

In Figure \ref{visc} we plot $\frac{1}{\eta}$ as a function of the system linear size, $\frac{1}{\rho^{\frac{1}{3}}}$ (here $\rho$ is liquid density), for liquid Ar along the adiabat and observe a quadratic dependence. This implies $\omega^{\rm bare}_{\rm F}\propto a^2$. Together with $\omega_{\rm F}\propto\frac{1}{a}$ due to the redshift, this gives $\omega_{\rm F}\propto a$.

\begin{figure}
\begin{center}
{\scalebox{0.4}{\includegraphics{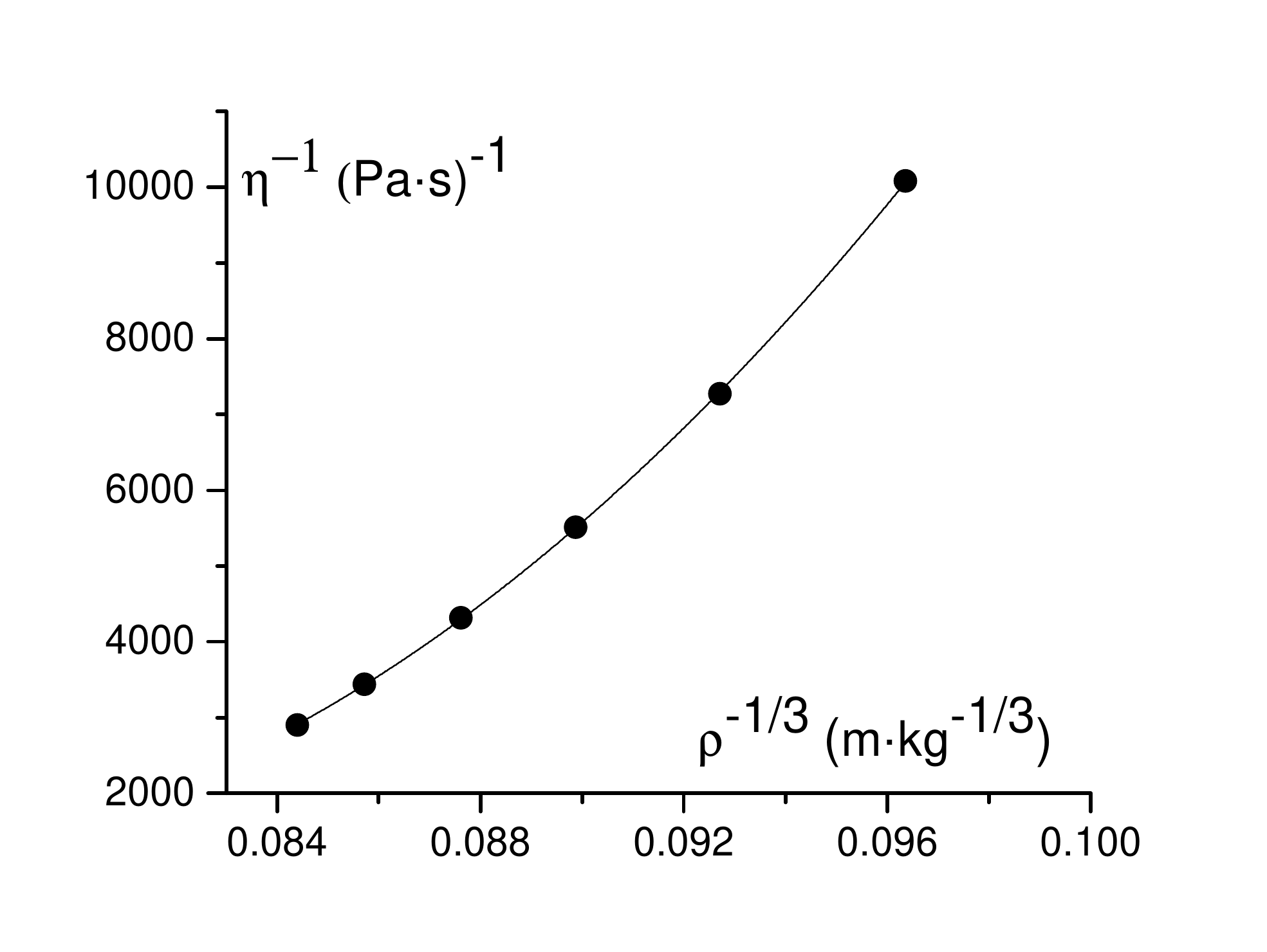}}}
\end{center}
\caption{Inverse of viscosity versus $\frac{1}{\rho^{\frac{1}{3}}}$ for liquid Ar along the adiabat at constant entropy $S=80$ J/(mol$\cdot$K). The data are from \cite{nist}. The line shows the quadratic fit ($R^2=0.99996$).}
\label{visc}
\end{figure}

Recalling that the vacuum energy density is given by (\ref{zero}) and using $\omega_{\rm F}\propto a$, we re-write (\ref{zero}) as

\begin{equation}
\rho_\Lambda=\rho_0\left(1-\left(\frac{a}{a_{\rm max}}\right)^{m}\right)
\label{1}
\end{equation}

\noindent where $m=n+1$.

In (\ref{1}), $\rho_0$ is the vacuum energy density at small $a$ ($a\ll a_{\rm max}$) or small $\omega_{\rm F}$, corresponding to the field mostly oscillating in one well and rare jumps between different wells. The increase of $a$ to its maximal value $a_{\rm max}$ gives $\rho_\Lambda=0$, corresponding to $E_0=0$ at $\omega_{\rm F}=\omega_{\rm max}$ in (\ref{zero}). As in (\ref{zero}), $\rho_\Lambda$ remains 0 for $a>a_{\rm max}$.

We note that varying energy components of the cosmological field were discussed before (see, e.g., Refs. \cite{sahni,stein,carvalo,basilakos,sola2} and references therein). In these discussions, the dynamical component is postulated in order to account for the experimental data, or assumptions are made about dynamics. Here, we find that varying vacuum energy is a natural consequence of evolving field modes as a result of self-interaction, as is the case for liquid modes. This assertion follows once a generic double-well form of the interaction potential is assumed. We also note that zero vacuum energy emerges in other approaches due to cancellation effects (see, e.g., Refs. \cite{carroll,klinkhamer}).

Using (\ref{1}) in the equation for energy conservation, $\dot{\rho}_\Lambda+3\frac{\dot{a}}{a}\left(\rho_\Lambda+P_\Lambda\right)=0$, gives $3P_\Lambda=\rho_0m-\rho_\Lambda(m+3)$ (see \cite{fn}), and Eq. (\ref{2}) becomes

\begin{equation}
\frac{\ddot{a}}{a}=-\frac{4\pi G\rho_0}{3c^2}\left((m+2)\left(\frac{a}{a_{\rm max}}\right)^m-2\right)
\label{3}
\end{equation}
\noindent

For $a\ll a_{\rm max}$, the right-hand side term in Eq. (\ref{3}) is nearly constant and positive, giving positive $\ddot{a}$ and inflating de Sitter Universe. As $a$ approaches $a_{\rm max}$, the right-hand side term in (\ref{3}) and $\ddot{a}$ become negative. This is the result of pressure $P_\Lambda$ becoming positive due to the reduction of the vacuum energy when $a$ approaches $a_{\rm max}$ (see (\ref{1})). The dynamics of $a$ has a stable equilibrium point at $a=a_{\rm eq}=a_{\rm max}\left(\frac{2}{m+2}\right)^{\frac{1}{m}}$ where $\ddot{a}=0$. The existence of the equilibrium point and sign variability of $\ddot{a}$ holds for any number of field components $n$.

Below we consider a single-component scalar field $\phi$ discussed in inflationary theories \cite{liddle}. For $n=1$ ($m=2$), (\ref{1}) and (\ref{3}) read:

\begin{equation}
\rho_\Lambda=\rho_0\left(1-\left(\frac{a}{a_{\rm max}}\right)^{2}\right)
\label{11}
\end{equation}

\begin{equation}
\frac{\ddot{a}}{a}=-\frac{8\pi G\rho_0}{3c^2}\left(2\left(\frac{a}{a_{\rm max}}\right)^2-1\right)
\label{31}
\end{equation}
\noindent

The equilibrium point $a_{\rm eq}$ at which $\ddot{a}=0$ in (\ref{31}) is

\begin{equation}
a_{\rm eq}=\frac{a_{\rm max}}{\sqrt{2}}
\label{32}
\end{equation}

We re-write (\ref{31}) as

\begin{equation}
\ddot{a}=Ba-Ca^3
\label{4}
\end{equation}
\noindent

\noindent with positive constants $B=\frac{8\pi}{3}\frac{G\rho_0}{c^2}$ and $C=\frac{16\pi}{3}\frac{G\rho_0}{c^2a_{\rm max}^2}$.

We recognize (\ref{4}) as the Duffing equation discussed in a non-linear theory as a common case
study. The equation with arbitrary $B$ and $C$ is solvable in terms of Jacobi elliptic functions $\sn(t,k)$, $\cn(t,k)$ or $\dn(t,k)$, where $k$ is the modulus of the elliptic integral \cite{kosevich,ryzhik}.

Eq. (\ref{4}) corresponds to the motion in an effective potential function $U^{\rm eff}(a)$, where $\ddot{a}=-\frac{dU^{\rm eff}(a)}{da}:$

\begin{equation}
U^{\rm eff}(a)=-\frac{Ba^2}{2}+\frac{Ca^4}{4}
\label{5}
\end{equation}
\noindent

$U^{\rm eff}(a)$ has a double-well form similar in shape to the potential in Fig. \ref{wells}. Since $a$ is positive, the physical solution at small enough energy corresponds to the periodic motion in the right well of $U^{\rm eff}(a)$ where $0<a<\sqrt{\frac{2B}{C}}=a_{\rm max}$. The solution is given by the delta amplitude elliptic function $\dn$:

\begin{equation}
a=R\dn(\omega t,k)
\end{equation}
\noindent where $R$, frequency $\omega$ and modulus $k$ depend on parameters $B$, $C$ and system's energy \cite{kosevich}.

\section{Cosmological implications}

As follows from (\ref{31}), there are two parameters in the model: $\rho_0$ and $a_{\rm max}$. More information about their range will enable quantitative evaluations of quantities of interest such as $R$, $\omega$ and other properties. However, some qualitative insights can be discussed already.

The proposed picture predicts the oscillatory motion of $a$. Hence, the current expansion with positive $\ddot{a}$ is predicted to be followed by $\ddot{a}<0$ and subsequent decrease of $a$. We note that direct differentiation shows that for an elliptic function such as $\dn(t,k)$, there is always a range of arguments where both $\dot{a}$ and $\ddot{a}$ are positive, as is currently observed.

An appealing feature of an oscillatory model is that, contrary to other inflationary scenarios, it is not faced with issues related to the beginning of space and time, predicting the future of the Universe or invoking an antropic principle \cite{roos}. We note that the oscillatory models of the Universe were discussed before (see, e.g. \cite{stein1,mulryne} and references therein), albeit with different underlying mechanisms. A distinct feature of the proposed model is that it does not need to involve very small length scales such as Planck or string sizes where yet unknown physics operates. The minimal value of $a$, $a_{\rm min}$, depends on the energy of the system oscillating in (\ref{5}) and can be large.


The expanding phase of the oscillatory motion with $\ddot{a}>0$ brings about the same benefits of inflation such as commonly discussed flatness and isotropy \cite{roos,liddle,linde}. The minimal value of $a$, $a_{\rm min}$, depends on the energy of the system oscillating in (\ref{5}). For small enough energy, $\frac{a_{\rm eq}}{a_{\rm min}}$ is large during the inflation phase.

The oscillatory behavior in this model is due to the variation of $\rho_\Lambda(a)$ only and does not involve mass density $\rho_m(a)$. $\rho_m(a)\propto\frac{1}{a^3}$ can be added to the right side of the acceleration equation (\ref{2}) but its contribution is expected to be small compared to $\rho_\Lambda(a)$ in (\ref{1}) or (\ref{11}) for large $a$. We note that $\rho_m(a)$ and $\rho_\Lambda(a)$ do not change independently because both parametrically depend on $a$. However, we do not expect to find the commonly discussed relationship between $\rho_\Lambda$ and $\rho_m$ \cite{carroll} because it involves the assumption of constant $\rho_\Lambda$.

Another implication is related to the interpretation of discrepancy between small experimental $\ddot{a}$ and large vacuum energy estimated from the field theory, the discrepancy discussed in the cosmological constant problem. If, as is often assumed, $\rho_\Lambda$ is constant, $\rho_\Lambda=\rho_0$, the acceleration equation, $\frac{\ddot{a}}{a}=-\frac{8\pi G}{3}\rho_0$, implies that large $\rho_0$ gives large $\ddot{a}$ \cite{carroll}. This is not the case if $\rho_\Lambda$ is variable: Eq. (\ref{31}) shows that $\ddot{a}$ can be small even if $\rho_\Lambda$ is large. Indeed, $\ddot{a}$ in (\ref{31}) remains small as long as $a$ is close to its equilibrium value $a_{\rm eq}$ in (\ref{32}) even though $\rho_\Lambda$ is large and close to $\rho_0$ in (\ref{11}) at and around $a_{\rm eq}$.

\section{Summary}

We proposed a non-perturbative approach to strong field interactions based on quasiparticles. This results in a variable vacuum energy that depends on the state of the system. If the field is related to inflation, the equation for $a$ has a stable equilibrium point, predicting the oscillatory behavior of $a$ and other cosmological effects.

I am grateful to D. Mulryne and V. V. Brazhkin for discussions and to Royal Society for support.


\begin{thebibliography}{99}
\bibitem{fn} For small $a$, $a\ll a_{\rm max}$, $\rho_\Lambda=\rho_0$ from (\ref{1}), giving $P_\Lambda=-\rho_\Lambda$ and Lorentz invariance. The latter relationship does not apply when $a$ approaches its boundary $a_{\rm max}$.
\end{thebibliography}
\end{document}